\documentclass[showpacs,twocolumn,prl]{revtex4}
\usepackage{graphicx}
\newcommand{\be}{\begin{equation}}
\newcommand{\ee}{\end{equation}}
\newcommand{\bea}{\begin{eqnarray}}
\newcommand{\eea}{\end{eqnarray}}

\newcommand{\la}{\langle}
\newcommand{\ra}{\rangle}

\newcommand{\bwt}{\begin{widetext}}
\newcommand{\ewt}{\end{widetext}}
\begin{document}
\title{Quantum mechanics of spin transfer in coupled electron-spin chains}
\author{Wonkee Kim$^{1}$, L. Covaci$^{1,2}$, F. Dogan$^{1}$, and F. Marsiglio$^{1}$}
\affiliation{
$^1$Department of Physics, University of Alberta,
Edmonton, Alberta, Canada, T6G~2J1
\\
$^{2}$Department of Physics and Astronomy, University of British
Columbia, Vancouver, British Columbia, Canada V6T 1Z1 }

\begin{abstract}

The manner in which spin-polarized electrons
interact with a magnetized thin film is currently described
by a semi-classical approach. This in turn provides our present
understanding of the spin transfer, or spin torque phenomenon.
However, spin is an intrinsically quantum mechanical quantity.
Here, we make the first strides towards a
fully quantum mechanical description of spin transfer through spin
currents interacting with a Heisenberg-coupled spin chain.
Because of quantum entanglement, this requires a 
formalism based on the density matrix approach.
Our description illustrates how individual spins in the chain
time-evolve as a result of spin transfer.

\end{abstract}


\date{\today}
\maketitle

The problem of switching the spin state of a ferromagnet has its
origins in magnetization reversal through an applied magnetic field
\cite{choi01}.
The present understanding of this process is
based on the (numerical) solution of the so-called
Landau-Lifshitz-Gilbert (LLG) equations \cite{mansuripur88}. These
equations define a space-dependent magnetization vector, whose
magnitude at each spatial coordinate is held fixed, but whose
direction is determined by a (classical) dynamical equation which
relates the rate of change of magnetization to the applied torque.
The applied torque is determined by an effective magnetic field,
which represents a mean field as experienced by the magnetization at
that particular spatial coordinate. A damping term is also required,
which is crucial for the magnetization reversal to take place.
%%%%%%%%%%%%%%%%%%%
A semi-classical approach \cite{slonczewski96,berger96}
was adopted to describe spin transfer \cite{myers99} from a
spin-polarized current to a ferromagnet.
The description of this phenomenon
requires a quantum mechanical treatment of the spin current, while a
classical notion of the magnetic moment
\cite{bazaliy98,waintal00,kim04}
suffices for a qualitative understanding of the spin torque process.
Most intriguing from our point of view is that a phenomenological
damping process is not required for spin torque to occur; in this
respect a spin current is profoundly different from an applied
magnetic field.

However, more recently, spin transfer has been recognized as
important, both in spintronics
\cite{myers99,weber01,wolf01,kiselev03,krivorotov05,hirjbehedin06},
and in quantum information\cite{nielsen00,leuenberger01}. For these
reasons, and because, after all, spin is an intrinsically quantum
property, it is critical to understand spin transfer as a fully
quantum mechanical process. This way, even within the more
conventional applications where the LLG approach has been fairly
successful, one can understand and perhaps push the limitations of
the semi-classical approach.
In this paper we construct the simplest but most important
elements of a model
with which one can illustrate the spin dynamics quantum mechanically.
Consider an electron
incident on a ferromagnetic chain with $N_{s}$ coupled local spins,
all prepared magnetized in a particular direction. Unless the
electron spin is parallel to the local spins, spin will be
transferred to the chain from the electron. In the semi-classical
picture, each spin $({\bf S})$ is considered as a classical vector;
in particular the magnitude of the vector remains constant, while
its direction is allowed to change. On the other hand, within a
quantum mechanical description, ${\bf S}$ is an operator
\cite{avishai01,kim05}. It is $\la {\bf S}^{2}\ra$ that remains
constant; in contrast, $\la {\bf S}\ra$ changes during the transfer
process, and can even vanish momentarily.

To describe complete spin transfer requires a current of polarized
electrons. This introduces another purely quantum mechanical concept
associated with spin transfer: entanglement between an electron and
the localized spin chain. Even after the electron moves away from
the spin chain, the quantum states of the electron and localized
spins are entangled in spin space. Thus, to illustrate the dynamics
of the chain we should know how to describe a subsystem (spin chain)
of a composite system (spin chain and electron). In particular, if
we send another electron to interact with the chain after a
preceding electron leaves the chain, it is crucial to decouple the
spin chain from the first electron quantum mechanically. 
The density matrix formalism \cite{feynman72} is the only way
to serve this purpose.
Suppose a system consists
of two subsystems A and B. This composite system is assumed to be
closed. Initially, A and B are separated and no interaction takes
place between the two. In other words, A and B are further apart
than the interaction range of the two. The quantum states of A and B
would be $|\psi_{A}\ra$ and $|\psi_{B}\ra$, respectively. Imagine,
now, the two subsystems come together and interact with one another.
For example, in our case, the mobile subsystem A (the electron)
moves into the region of interaction with the stationary subsystem B
(the localized spin chain), they interact, and then A leaves the
domain of B. Let us assume the interaction lasts for a finite amount
of time, say $t_{0}$. The dynamics of the composite system will be
described by the total Hamiltonian $H = H_{A} + H_{B} + H_{int}$,
where $H_{A}$ and $H_{B}$ are the Hamiltonians of A and B,
respectively, while $H_{int}$ is the interaction between the two.

The density matrix $\rho^{tot}$ of the total system at $t=0$ is
$\rho^{tot}(0) = \rho^{A}(0)\otimes\rho^{B}(0)$, where
$\rho^{A}(0)=|\psi_{A}\ra\la \psi_{A}|$ and
$\rho^{B}(0)=|\psi_{B}\ra\la \psi_{B}|$. The quantum states of A and
B are assumed to be pure initially, which means that these states
are exactly known at $t=0$. The time evolution of $\rho^{tot}$ is
given by 
\be 
\rho^{tot}(t) = U(t)\rho^{tot}(0)U^{\dagger}(t) 
\ee
where $U(t) = e^{-iHt}$. The expectation value of an operator is
calculated as follows: \be \la\Omega\ra =
\mbox{Tr}_{A,B}\left[\Omega\rho^{tot}(t)\right] =\sum_{a,b}\la a|\la
b|\Omega\rho^{tot}(t)|b\ra|a\ra \ee where $\{|a\ra\}$ and
$\{|b\ra\}$ are orthonormal sets associated with the subsystems A
and B, respectively. In particular, for an operator $\Omega_{B}$ of
B, the expectation value is $\la\Omega_{B}\ra =
\mbox{Tr}_{B}\left[\Omega_{B}\rho^{B}\right]$ where the reduced
density matrix of the subsystem B, $\rho^{B}$, is obtained by
tracing out the states of the subsystem A: 
$\rho^{B} = \mbox{Tr}_{A}\left[\rho^{tot}(t)\right]$.
While this relation is obvious, it serves to indicate the
significance of the reduced density matrix $\rho^{B}$. After the two
subsystems no longer interact at $t=t_{0}$, the quantum states of B
are represented solely by $\rho^{B}$. The quantum state of B is not
pure but mixed; in other words, it is impossible to represent
$\rho^{B}(t_{0})$ as $|\Psi_{B}\ra\la\Psi_{B}|$, where
$|\Psi_{B}\ra$ is a quantum state exactly known at $t_{0}$. It is a
generic property of quantum entanglement that a subsystem is in
mixed states after getting decoupled even if the composite state is
pure\cite{nielsen00}.
If another subsystem (second electron) is introduced and interacts
with B at time $t_{0}$, $\rho^{B}(t_{0})$ should be used as a part of the
initial density matrix corresponding to subsystem B. Subsystem A can
be introduced as for the first electron, but representing the next
electron. Then the initial density matrix is
$\rho^{tot}(t_{0})=\rho^{A}(t_{0})\otimes\rho^{B}(t_{0})$, where
$\rho^{A}(t_{0})=\rho^{A}(0)$ by preparation, and its time evolution
will be $\rho^{tot}(t) = U(t)
\left[\rho^{A}(t_{0})\otimes\rho^{B}(t_{0})\right]U^{\dagger}(t)$
for $t > t_{0}$. We then apply the same procedure systematically in
this way to allow a beam of electrons to individually interact with
the spin chain.

As a model Hamiltonian \cite{kim06} for spin transfer from incident
electrons to a localized spin chain we use%
\be%
H=-\tilde{t} \sum_{<i,j>\sigma}c^{\dagger}_{i\sigma}c_{j\sigma}
-2\sum_{l=1}^{N_s}J_{0}{\bf\sigma}_{l}\cdot{\bf S}_{l}
-2\sum_{l=1}^{N_s-1}J_{1}{\bf S}_{l}\cdot{\bf S}_{l+1}%
\label{model_H}
\ee
where $c^{\dagger}_{i\sigma}$ creates an electron with spin $\sigma$
at site $i$, ${\bf S}_{l}$ is a localized spin operator at site $l$,
$\tilde{t}$ is a hopping amplitude between nearest neighbor sites,
so that the electron can move along the length of the entire chain,
and $J_{0}$ is the coupling between the electron spin and a local
spin. Note that $J_{0}$ coupling takes place only when an electron
is on the site of the local spin. The parameter $J_{1}$ is a
coupling between two neighboring localized spins. For a
ferromagnetic chain $J_{1}>0$. However, this formalism can be
applied for arbitrary values of $J_{0}$ and $J_{1}$; for example,
one can study spin transfer to an antiferromagnet \cite{nunez06}
using this model Hamiltonian. Moreover, these coupling parameters
can be site-dependent. A convenient notation is to represent $J_{0}$
as a vector with $N_{s}$ components and $J_{1}$ as a vector with
$N_{s}-1$ components whenever necessary.
Since we do not include longer range spin-spin
interactions, which would lead to domain formation, we can use a
site-dependent $J_1$ to mimic domains.
Scalar potentials are not taken into account because they are
irrelevant to spin transfer.

Our calculation scheme is as follows: we propagate electrons, one
after another, towards the ferromagnetic spin chain. These electrons
are spin-polarized in the $+z$ direction, while, initially, all
local spins are aligned (for example, along the $x$, or the $-z$
direction). Depending on the momentum of the incident electron, a
time $t_0$ can be determined, after which the electron and the spins
no longer interact. At $t_{0}$, the reduced density matrix of the
spin chain is evaluated from the total density matrix by decoupling
the spins from the electron. At the same time we introduce another
electron to interact with the spins. This reduced spin density
matrix is used to construct an initial density matrix of the total
system (spin chain and new electron). This process is repeated until
eventually all local spins become aligned with the incident electron
spin. 

The density matrix of the total system (spin chain and electron) is
$\rho^{tot}(0) = \rho^{S}(0)\otimes\rho^{el}(0)$ at $t=0$, where
$\rho^{el}(0) = |\phi_{0}\ra\la\phi_{0}|$. The initial state of the
electron is defined as $|\phi_{0}\ra = \sum_{i}\varphi_{i}|i,+\ra$,
where the amplitudes $\varphi_{i}$ are chosen to represent a Gaussian wave
packet with a mean momentum $k_{0}$ and a mean position $x_{0}$.
We choose
$k_{0}=\pi/2$, and $x_{0}$ sufficiently far from the chain such that
initially the wave packet is well defined outside the chain.
The spin state of the incident electron is $|+\ra$, i.e. along the $+z$
direction. 
The reduced density matrix of the spin chain can be
obtained by quantum-mechanically decoupling the chain from the
electron: $\rho^{S}(t)
=\sum_{i,\sigma}\la i,\sigma|\rho^{tot}(t)|i,\sigma\ra$, where
$|i,\sigma\ra$ is a state vector of an electron with spin $\sigma$
at a site $i$. 
As mentioned, the time evolution of $\rho^{tot}$ is given by
$\rho^{tot}(t) = U\rho^{tot}(0)U^{\dagger}$, where
$U = \exp[-iHt]$.
At $t_{0}$ we introduce another electron to interact with the chain.
Now the initial density matrix of the spin chain
and this electron is $\rho^{tot}(t_{0})=
\rho^{S}(t_{0})\otimes\rho^{el}(0)$, and its time evolution is
$\rho^{tot}(t>t_{0})=U\rho^{S}(t_{0})\otimes\rho^{el}(0)U^{\dagger}$.
After the $(N+1)$th electron, we obtain
\be
\rho^{S}_{N+1}(t) = \sum_{i_{N},\sigma_{N}} \la
i_{N},\sigma_{N}|U \rho^{tot}_{N}(Nt_{0})
U^{\dagger}|i_{N},\sigma_{N}\ra \;.
\ee
In our numerical calculation, we consider a ferromagnetic chain in a
state $|F\ra$ with $N_{s}$ local spins; thus,
$\rho^{S}(0)=|F\ra\la F|$. However, any initial spin configuration
can be assumed.
The expectation value of a spin
operator, say $S_{1z}$, is evaluated by $\la S_{1z}\ra =
\mbox{Tr}_{S}\left[S_{1z}\;\rho^{S}\right]$.

%\bwt
\begin{figure}[tp]
\begin{center}
\includegraphics[height=3.5in,width=3.in,angle=-90]{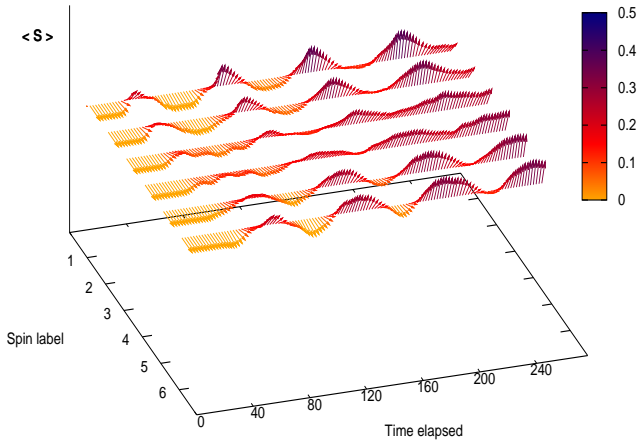}
\caption{
Time evolution of $6$ spins interacting with $4$ incoming electrons
one by one. The site-independent electron-spin coupling is
$J_{0} = 2$
and the ferromagnetic coupling is $J_{1} = 0.4$, which is also
site-independent. The arrows depict $\la{\bf S}\ra$
as three-dimensional vectors
for spatial as well as temporal behavior of individual spins, and
$\la S_{z}\ra$ is further visualized by colors.
}
\end{center}
\end{figure}
%\ewt

In order to describe spin transfer from spin-polarized incoming
electrons to a spin chain, we show dynamics of the spins in the chain.
We consider $6$ spins with $S=1/2$ initially pointing to the $x$ direction.
For the parameters we choose, the electron wave packet
no longer interacts with the chain at $t_{0} = 60$ (in dimensionless units).
Fig.~1 illustrates how the $6$ spins time-evolve while interacting with
$4$ incoming electrons one after another. The site-independent
electron-spin coupling is
$J_{0} = 2$ and the ferromagnetic coupling is
$J_{1} = 0.4$, which is also site-independent, in units of the hopping
parameter $\tilde{t} (\equiv 1)$.
The arrows depict $\la{\bf S}\ra$ as three-dimensional vectors
for spatial as well as temporal behavior of individual spins, and
$\la S_{z}\ra$ is further visualized by colors.
As discussed in Ref.\cite{kim05},
the coupling $J_0$ determines the degree of spin transfer, which depends
on $J_0$ non-monotonically.
Fig.~1 shows two distinct dynamical regimes. First, while an
electron wave packet is striking the spin chain, the $J_0$
interaction term in $H$ causes spin transfer,
and therefore spin flip. This will occur with every member of the spins
with multiple reflections of the electron within the chain included.
Once the electron has left
the spin chain, however, the spins continue to evolve in time, but
solely due to the $J_1$ term. The result is the
oscillating behavior depending on the value of $J_1$.

%\bwt
\begin{figure}[tp]
\begin{center}
\includegraphics[height=3.5in,width=3.in,angle=-90]{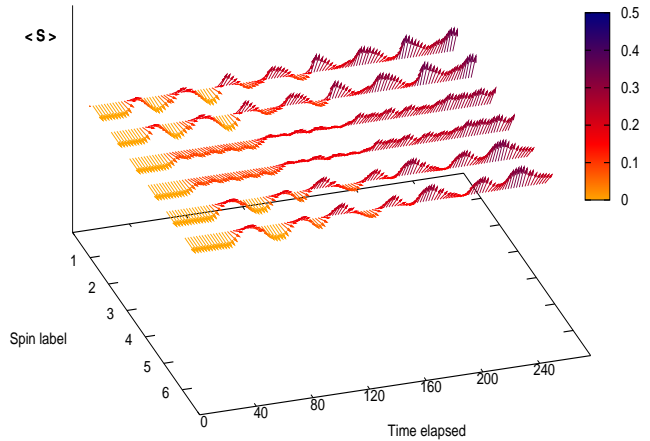}
\caption{
Time evolution of $6$ spins with a domain structure formed by $J_{1}$
interacting with $4$ incoming electrons one by one.
The coupling $J_{0}$ remains unchanged as in Fig.~1 while
the site-dependent $J_1 = (5,\;0.4,\;5,\;0.4,\;5)$ for $3$ domains.
}
\end{center}
\end{figure}
%\ewt

As mentioned earlier, parameterizing components of $J_1$,
one can envision a domain-structured spin chain. In Fig.~2,
the coupling $J_{0}$ remains unchanged as in
Fig.~1 while we choose a site-dependent
$J_1 = (5,\;0.4,\;5,\;0.4,\;5)$ for $3$ domains to form.
The dynamics of spins is very different from the previous case.
Because of the strong coupling between three spin pairs, two spins
in each pair evolve simultaneously. The first and the last
spin pairs oscillate against each other while the middle pair
evolves almost monotonically. This overall behavior of three spin pairs
is qualitatively the same as one of weakly coupled three local spins.
Note that as more electrons are sent very little changes; for these
parameters, five electrons are nearly sufficient to align the
entire chain along the $+z$ direction.

Another issue we wish to address in this paper concerns the excitations in the spin chain
induced by the spin transfer process. We do not include a magnetic field in the
Hamiltonian Eq.~(\ref{model_H}) because we investigate excitations
only due to the relative spin configuration in the spin chain. If a
magnetic field is included, additional excitations will be induced
depending on the orientation of local spins with respect to the
field. For investigation of the induced excitations, we consider
only site-independent couplings $J_{0}$ and $J_{1}$.
The excitations induced by spin transfer are, in fact, associated
with the energy transfer between the incident electrons and the spin
chain.
The energy of the chain is given by the expectation value
$E_{S}$ of the Hamiltonian of the chain $H_{S}=-2J_{1}\sum_{l}{\bf
S}_{l}\cdot{\bf S}_{l+1}$; namely, $E_{S} =
\mbox{Tr}_{S}\left[H_{S}\rho^{S}\right]$. 
We demonstrate that $E_{S}$ depends on the coupling parameter $J_{1}$,
and energy is not monotonically transferred to the chain 
from incoming electrons.
As local spins
start to align with one another in the direction of the incident
electron spin, energy would be transferred back to incoming
electrons from the chain as illustrated in Fig.~3.
This backwards energy transfer can be understood in a sense
that excitation and de-excitation
are relative in the absence of a magnetic field.
Fig.~3 is for
a spin chain interacting with $25$ incident electrons
one by one. We set $J_{0}=1$ but choose various values of $J_{1}$.
The mean momentum $k_{0}$ of incident electrons is $\pi/2$ in the
main frame while it varies in the inset to show that the energy
transfer also depends on the momentum of the incident electrons. If
$J_{1}$ is large enough, say $J_{1}>5$, then no excitations
occur
and $E_{S}/E_{S}(0) = -1$, where $E_{S}(0)=J_{1}\left(N_{s}-1\right)/2$
for a chain with
$N_{S}$ local spins. Since there is no preferred direction due to
the absence of a magnetic field, $E_{S}$ remains unchanged as all
spins rotate in unison. This can be shown analytically.

\begin{figure}[tp]
\begin{center} \includegraphics[height=2.55in,width=3.in]{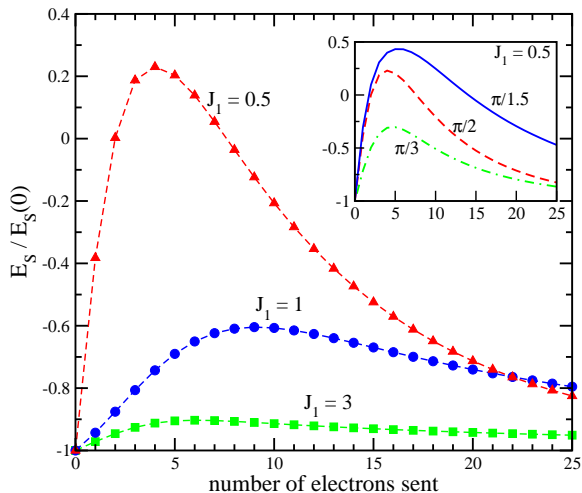}
\caption{
Excitation induced by spin transfer $E_{S}$ normalized by
$E_{S}(0)=J_{1}\left(N_{s}-1\right)/2$ as a function of the number of electrons
sent in. The spin chain interacts with $25$ incident electrons
with $J_{0}=1$ and various values of $J_{1}$.
The momentum $k_{0}$ of incident electrons is $\pi/2$ in the
main frame while it varies in the inset.
}
\end{center}
\end{figure}

In summary, we have outlined a quantum-mechanically rigorous
framework with which one can calculate the spin
transfer from a spin-polarized electron current to a spin chain.
A brute force method is possible, but for a progression of
incident electrons, this will quickly expand the Hilbert space to exceed
the reach of modern computers. Application of the density matrix formalism
not only solves this difficulty but also correctly deals with quantum
entanglement inherent to the spin transfer problem.
We have performed computations for a rather
simplistic system; nonetheless, included are two essential elements
of spin transfer; spins as quantum operators and entanglement.

As far as the physics of spin transfer is concerned, it is clear that the
semi-classical assumption that the magnitude of the magnetization
vector remains constant during spin transfer is generally violated.
In particular, when the ferromagnetic exchange coupling is weak we
find regimes where the magnitude of the spin vector can almost
vanish. How this might alter our understanding of experimental
observations will be the subject of further investigation.

This work was supported in part by the Natural Sciences
and Engineering Research Council of Canada (NSERC), by ICORE
(Alberta), and by the Canadian Institute for Advanced Research
(CIAR). Numerical calculations have been performed using computational
facilities of the Academic Information and Communication Technologies
(AICT) and the WestGrid. 
%W.K. and F.D. acknowledge M. Fujinaga at AICT for technical help.

\bibliographystyle{prl}

\end{document}